\documentclass{article}
\usepackage{graphicx}

\begin{document}

\title{OPERA Superluminal Neutrinos per Quantum Trajectories}

\author{Edward R. Floyd \\
10 Jamaica Village Road, Coronado, CA 92118-3208, USA \\
floyd@mailaps.org}

\date{\today}

\maketitle

\centerline{\itshape Dedicated to Lynne Floyd, 1 July 1934 -- 16 October 2011}

\bigskip

\begin{abstract}
Quantum trajectories are used to study OPERA findings regarding superluminal neutrinos.  As the applicable stationary quantum Klein-Gordon equation is real,
real quantum reduced actions and subsequent real quantum trajectories follow.  The requirements for superluminal neutrinos are examined.  A neutrino that is
self-entangled by its own backscatter is shown to have a nonlocal quantum trajectory that may generate a superluminal transit time.  Various cases are shown to produce theoretical superluminal neutrinos consistent with OPERA neutrinos.  Quantum trajectories are also shown to provide insight into neutrino oscillations.
\end{abstract}

\bigskip

\small


\bigskip


\normalsize

\bigskip

\section{Introduction}

The OPERA collaboration has initially reported, with due caution, that superluminal neutrinos have been observed [\ref{bib:opera}].
Due to the profoundness of this event, the initial OPERA report has suggested that this experiment should be replicated for confirmation.
The OPERA collaboration also intentionally offered neither theoretical nor phenomenological explanations.  The physics community has responded
with a plethora of papers addressing the causes for superluminal neutrinos including three papers by Matone [\ref{bib:m},\ref{bib:m2}] and
Faraggi [\ref{bib:f}], who have couched their work in the quantum equivalence principle (QEP) [\ref{bib:pl450}--\ref{bib:bfm}].
They have developed real quantum trajectories for neutrinos. [\ref{bib:m}--\ref{bib:f}].  QEP and the quantum trajectory representation are mutually
compatible and are embedded in a common quantum Hamilton-Jacobi formulation.

Herein, I develop real quantum trajectories for a neutrino that is self-entangled by its own backscatter and discuss its implications bearing upon
the OPERA experiment.  ``Backscatter" herein means that each neutrino contains an internal degree of backscatter, which is discussed further in \S2.
Two non-independent quantum effects are shown to induce superluminal propagation: self entanglement, which is the entanglement among the spectral components (one of which is backscatter) within an individual neutrino; and nonlocality.  Possible physical causes of backscatter are discussed.  Backscatter is shown to be a possible cause of neutrino oscillations.  This relativistic quantum-trajectory investigation is embedded in a quantum Hamilton-Jacobi formulation of the associated stationary Klein-Gordon equation (SKGE).

Some in the physics community have investigated OPERA superluminal neutrinos as a group velocity phenomenon [\ref{bib:mb}--\ref{bib:tm}].
The study of group velocities is in the $\psi$-representation in Hilbert space and has much in common with the quantum trajectories representation.
But quantum trajectories are presented in configuration space plus time consistent with the underlying quantum Hamilton-Jacobi formulation.
Quantum trajectories are not burdened with issues of wave-packet integrity including the case of strongly peaked spectral components in the
wave-number domain of a particle that, if treated as distinct entities, would spatially separate widely into distinct entities whose overlap
and entanglement with each other would rapidly decrease [\ref{bib:fp37a}].  A quantum trajectory is derived from the quantum reduced
action (a generator of quantum motion) for the particle as a whole.  The quantum reduced action contains the internal backscatter so that all entanglement is
retained.  This entanglement induces nonlocal motion in the subsequent quantum trajectory.

Section 2 develops the formulation for relativistic quantum trajectories for neutrinos with internal backscatter.  The subsequent quantum trajectories give insight into neutrino oscillation. Backscatter is identified as a spectral component of a dichromatic $\psi$ that travels in the counter direction of $\psi$'s other spectral component. In  \S3, quantum trajectories show that while a set of the constants of quantum motion for superluminal transit times for the OPERA experiment always exist, other sets of constants of the quantum motion result in subluminal transit times.  For superluminal neutrinos, it is shown that the OPERA experiment must favorably bias the selection of constants of quantum motion.  Some favorable systematic biases are examined.  Section 4 presents the conclusion.

\section{Formulation}

Matone [\ref{bib:m}] and Faraggi [\ref{bib:f}] have shown that the applicable relativistic quantum stationary Hamilton-Jacobi equation (RQSHJE)
for the OPERA neutrinos has the same form as the nonrelativistic quantum stationary Hamilton-Jacobi equation (NQSHJE).  The RQSHJE for the
OPERA experiment may be given in one-dimension, $q$, by [\ref{bib:f},\ref{bib:ijtp27}]

\begin{equation}
\left(\frac{\partial W}{\partial q}\right)^2 + m^2c^2 - \frac{E^2}{c^2} + \frac{\hbar^2}{2}\langle W;q \rangle = 0
\label{eq:rqshje}
\end{equation}

\noindent where $W$ is the relativistic quantum reduced action (Hamilton's quantum characteristic function), which is a generator of the quantum motion.
And $\langle W;q \rangle$ is the Schwarzian derivative of $W$ with respect to $q$ and given explicitly by

\[
\langle W;q \rangle = \frac{\partial^3 W}{\partial q^3} - \frac{3}{2}\left(\frac{\partial^2 W}{\partial q^2}\right)^2.
\]

\noindent  The Schwarzian derivative contains the quantum effects and makes the RQSHJE a third-order non-linear differential equation that requires
more constants of the quantum motion than the relativistic classical stationary Hamilton-Jacobi equation, which is a first-order nonlinear differential
equation [\ref{bib:prd29}].

A general solution for $W$ of Eq.\ (\ref{eq:rqshje}) is given within an integration constant by
[\ref{bib:m}--\ref{bib:f},\ref{bib:fm},\ref{bib:prd34}--\ref{bib:pe5}]

\begin{equation}
W(q) = \hbar \arctan\left(\frac{A \theta(q) + B \hat{\theta}(q)}{C \theta(q) + D \hat{\theta}(q)}\right)
\label{eq:gw}
\end{equation}

\noindent where $\{\theta,\hat{\theta}\}$ is a set of independent, real solutions of the associated one-dimensional SKGE,

\[
-\hbar^2 c^2 \frac{\partial^2 \psi}{\partial q^2} + (m^2c^4 - E^2) \psi = 0.
\]

\noindent  The coefficients $(A,B,C,D)$ are real constants here.  The Wronskian ${\mathcal W}$ is normalized so that

\begin{equation}
{\mathcal W}^2(A \theta + B \hat{\theta},C \theta + D \hat{\theta}) = (AD-BC) {\mathcal W}^2(\theta,\hat{\theta}) = (\hbar c)^{-2}.
\label{eq:wn}
\end{equation}

\noindent The formula $(\hbar c)^{-2}$ of the Wronskian normalization for the relativistic case replaces the analogous formula $2m/\hbar^2$ of the Wronskian normalization for nonrelativistic case [\ref{bib:prd34},\ref{bib:hm},\ref{bib:milne}].  (The quantum trajectory representation of quantum mechanics uses a Wronskian normalization while the $\psi$-representation uses a Born probability normalization.)   The coefficients $(A,B,C,D)$ are specified by the normalization of the Wronskian and the
initial conditions for $W$ analogous to those for the NQSHJE, [\ref{bib:pl450},\ref{bib:fm},\ref{bib:prd29},\ref{bib:prd34}].
The coefficients also must obey $AD-BC \ne 0$, otherwise by the principle of superposition $A \theta(q) + B \hat{\theta}(q)$
and $C \theta(q) + D \hat{\theta}(q)$ would be redundant and consequently $W(q)$ would be constant which is forbidden [\ref{bib:pl450},\ref{bib:fm}].
The general solutions of the RQSHJE and the SKGE imply each other as has been shown elsewhere for the analogous NQSHJE and
Schr\"{o}dinger equation [\ref{bib:pl450},\ref{bib:fpl9}].  One can always find a set $\{\vartheta,\hat{\vartheta}\}$ of independent solutions
for the SKGE for which the coefficients $B,C=0$ by setting $\vartheta=A\theta+B\hat{\theta}$ and $\hat{\vartheta}=C\theta+D\hat{\theta}$.  Thus,
by superposition the quantum reduced action, Eq.\ (\ref{eq:gw}), may alternatively be expressed as $W = \hbar \arctan[\vartheta(q)/\hat{\vartheta}(q)]$.

In the course of this exposition, a redundant set of $\theta$s is nevertheless used to facilitate insight into the OPERA quantum trajectories.
Let us consider that the particular quantum reduced action for OPERA neutrinos be specified by

\begin{equation}
W  =  \hbar \arctan{ \left( \frac{\alpha \sin(+kq) + \beta \sin(-kq - \varphi)}{\alpha \cos(+kq) + \beta \cos(-kq - \varphi)}\right)}
\label{eq:eprw}
\end{equation}

\noindent where $k = (E^2 - m^2c^4)^{1/2}/(\hbar c)$ and $\alpha^2 + \beta^2 = 1$ where unless explicitly stated otherwise $\alpha \ge \beta \ge 0$. The set
of redundant solutions to the SKGE is $\{\Gamma \sin(+kq),\Gamma \cos(+kq),\Gamma \sin(-kq-\varphi),\Gamma \cos(-kq-\varphi)\}$ where the $\Gamma$ is the Wronskian normalization factor given by $\Gamma = [(E^2-m^2c^4)(\alpha^2-\beta^2)]^{-1/2} = [\hbar k c(\alpha^2-\beta^2)]^{-1/2}$ consistent with Eq.\ (\ref{eq:wn}). The factor $\Gamma$ cancels out in Eq.\ (\ref{eq:eprw}).  The phase shift $\varphi$ is restricted to
$-\pi < \varphi \le \pi$.  The set $\{E,\beta,\varphi\}$ is a sufficient set of constants of the quantum motion to the describe the quantum reduced action $W(E,\beta,\varphi;q)$.  As $W$ does not explicitly appear in the RQSHJE while its derivatives of first three orders do, the conjugate momentum, $\partial W/\partial q$ is also a solution of the RQSHJE.  The conjugate momentum is given by

\[
\frac{\partial W}{\partial q} = \frac{\hbar k (\alpha^2-\beta^2)}{\alpha^2 + \beta^2 + 2\alpha\beta \cos(2kq+\varphi)}.
\]

\noindent  Note that $\partial W/\partial q \ne 0$ for $\alpha > \beta$ consistent with $W$ never being constant [\ref{bib:pl450},\ref{bib:fm}].   As developed elsewhere by a nonrelativistic analogy [\ref{bib:fp37a}], this particular form for $W$ and $\partial W/\partial q$ may be linked with an eigenfunction of the associated SKGE that is a Wronskian-normalized dichromatic wave function, $\psi$ whose spectral representation may be given by

\begin{eqnarray}
\psi & = & \overbrace{\Gamma [\alpha^2 + \beta^2 + 2\alpha \beta \cos(2kq+\varphi)]^{1/2}\exp \left[ i\arctan{ \left( \frac{\alpha \sin(+kq) + \beta \sin(-kq - \varphi)}{\alpha \cos(+kq) + \beta \cos(-kq - \varphi)}\right)} \right] }^{\mbox{eigenfunction of SKGE}} \nonumber \\
   & = & \underbrace{\Gamma \alpha \exp(+ikq)}_{\psi_{\mbox{\scriptsize forward}}} + \underbrace{\Gamma \beta \exp(-i\varphi) \exp(-ikq)}_{\psi_{\mbox{\scriptsize backscatter of SKGE}}}.
\label{eq:spectrum}
\end{eqnarray}

\noindent The intermediate formula in Eq.\ (\ref{eq:spectrum}) for $\psi$ labeled ``eigenfunction of SKGE" has an exponential argument that could be expressed as $iW/\hbar$ in the one-dimensional case.  The amplitude for the intermediate form for $\psi$ in Eq.\ (\ref{eq:spectrum}) is proportional to $(\partial W/\partial q)^{-1/2}$.  The final formula of Eq.\ (\ref{eq:spectrum}) exhibits the dichromatic resolution of $\psi$ into spectral components at $\pm k$.   The very existence of a dichromatic spectral resolution for $\psi$ is sufficient for entanglement as $\psi \ne \psi_{\mbox{\scriptsize forward}} \times \psi_{\mbox{\scriptsize backscatter}}$, which in turn induces nonlocality [\ref{bib:fpl9}].  This self-entanglement is also manifested in $W$, Eq.\ (\ref{eq:eprw}), as $W$ is not separable into two terms:
one containing only forward propagation and the other, only backscatter, that is $W \ne W_{\mbox{\scriptsize forward}}
+ W_{\mbox{\scriptsize backscatter}}$ [\ref{bib:fpl9}].  This particular selection of a redundant set of $\theta$s to describe $W$ lets one introduce a degree, $\beta$, of backscatter, $\{\Gamma \sin(-kq-\varphi),\Gamma \cos(-kq-\varphi)\}$ that coherently interferes in a self-entangled manner with the forward motion
$\{\Gamma \sin(+kq),\Gamma \cos(+kq)\}$ of amplitude $\alpha$ in the generator of quantum motion.  The quantum reduced action for the neutrino, Eq.\ (\ref{eq:eprw}), retains the self-entanglement due to backscatter no matter how large $q$ becomes.  The phase shift, $\varphi$ generalizes $W$ and $\psi$. If $\beta = 0$ (i.e., no backscatter), then $\alpha = 1$, $\Gamma = k^{-1/2}$, $W=+\hbar kq$ and $\psi= (\hbar k c)^{-1/2} \exp(+ikq)$, which would be rectilinear motion.

Jacobi's theorem is used to determine the relativistic equation of quantum
motion [\ref{bib:m}--\ref{bib:f},\ref{bib:fm},\ref{bib:prd34},\ref{bib:rc}]. As
Jacobi's theorem also determines the equation of motion in classical mechanics,
Jacobi's theorem does transcend across the division between classical and
quantum mechanics to give a universal equation of motion. Jacobi's theorem renders time generation as

\begin{equation}
\underbrace{t-\tau = \partial W/\partial E}_{\mbox{Jacobi's theorem}} =\underbrace{\overbrace{\frac{q}{c}}^{t_{c}}\,\times\,\overbrace{\frac{E}
{(E^2-m^2c^4)^{1/2}}}^{\mbox{relativistic factor},\ H_R}}_{\small t|_{\beta=0}}\,\times\,\underbrace{\frac{(\alpha^2-\beta^2)}{1 + 2 \alpha \beta
\cos(2kq+\varphi)}}_{\mbox{quantum factor},\ H_Q}
\label{eq:eom}
\end{equation}

\noindent where $\tau $ is the Hamilton-Jacobi constant coordinate (a nontrivial
constant of integration) that specifies the epoch and where $t_{c}$ is the time needed to transit the distance $q$ by light in a vacuum.
Let us set $\tau=0$ herein.  Interference between the spectral components (self-entanglement) of the associated $\psi$, Eq.\ (\ref{eq:spectrum}), is manifested
in the quantum trajectory by the cosine term in the denominator of Eq.\ (\ref{eq:eom}).  This self-entanglement is sustained in Eq.\ (\ref{eq:eom}) no matter
how large $q$ becomes.

The quantum factor, $H_Q(q)$, has extrema at

\begin{equation}
\frac{d(H_Q)}{dq} =  \frac{2k (\alpha^2 - \beta^2) \alpha\beta \sin(2kq+\varphi)}{[{1 + 2 \alpha \beta
\cos(2kq+\varphi)}]^2} = 0.
\label{eq:dfQdq}
\end{equation}

\noindent or

\begin{equation}
q = \frac{n\pi}{2k} - \frac{\varphi}{2k},\ \ \ n=0,\pm1,\pm2,\cdots.
\label{eq:qn}
\end{equation}

\noindent  The interference due to backscatter ($\beta \ne 0$) induces the quantum trajectories to oscillate between maxima and minima
times of propagation [\ref{bib:fp37a}].  For every $q$, there exist two $\varphi$s within the range $-\pi,+\pi$ that satisfy Eqs.\ (\ref{eq:dfQdq})
and (\ref{eq:qn}). One of which, $\varphi_{\mbox{\scriptsize max}}$, renders $\cos(2kq+\varphi_{\mbox{\scriptsize max}}) = -1$; the
other, $\varphi_{\mbox{\scriptsize min}}$, renders $\cos(2kq+\varphi_{\mbox{\scriptsize min}}) = +1$.  The different phase shifts, $\varphi_{\mbox{\scriptsize
max}}$ and $\varphi_{\mbox{\scriptsize min}}$, as constants of the quantum motion, specify different quantum trajectories on the $q,t$-plane that arrive
at the designated  $q$  at different times.

The local maxima propagation times, $t$, occur in the vicinity of the local maximum of $H_Q$ given by
$q = [(2n-1)\pi ]/(2k) - \varphi_{\mbox{\scriptsize max}}/(2k),\ n=0,\pm1,\pm2,\pm3,\cdots$; minima times at $q = n\pi /k - \varphi_{\mbox{\scriptsize min}}/(2k),\ n=0,\pm1,\pm2,\pm3,\cdots$.
 The quantum factor, $H_Q(q)$, has a maximum given by

\begin{equation}
H_{Q,{\mbox{\scriptsize max}}} = \frac{\alpha + \beta}{\alpha - \beta}, \ \ \ q = \frac{(2n-1)\pi}{2k} - \frac{\varphi_{\mbox{\scriptsize max}}}{2k}, \ \ \ n=0,\pm1,\pm2,\pm3,\cdots,
\label{eq:maxhq}
\end{equation}

\noindent and a minimum value given by

\begin{equation}
H_{Q,{\mbox{\scriptsize min}}}=\frac{\alpha - \beta}{\alpha + \beta}, \ \ \ q = \frac{n\pi}{k} - \frac{\varphi_{\mbox{\scriptsize min}}}{2k}, \ \ \ n=0,\pm1,\pm2,\pm3,\cdots.
\label{eq:minhq}
\end{equation}

\noindent The extrema, $H_{Q,{\mbox{\scriptsize max}}}$ and $H_{Q,{\mbox{\scriptsize min}}}$, form the envelope (caustics) for a wedge in
the $q,t$-plane with its apex at the origin.  This wedge is analogous to the light cone.  Note that $H_{Q,{\mbox{\scriptsize max}}}$
and $H_{Q,{\mbox{\scriptsize min}}}$ are functions of only the constant of the quantum motion $\beta$ for $\alpha = +(1-\beta^2)^{1/2}$.

The relativistic factor $H_R$ is a function of $E$ and $m$, but not of the variable $q$.  The relativistic factor for the OPERA experiment
for 17 GeV neutrinos may be approximated by Eq.\ (\ref{eq:eom}) as

\[
H_R = 1 + \frac{m^2c^4}{E^2} + O(m^4c^8/E^4)
\]

\noindent where $(m^4c^8/E^4) \ll 1$. For neutrinos, it is expected that $|1 - H_R| < 10^{-19}$ even in their most massive eigenstate [\ref{bib:opera}].
If $\beta$ is sufficiently large so that $H_{Q,\mbox{\scriptsize min}} \times H^{-1}_R < 1$, then the OPERA neutrino may be superluminal.  Herein,
$\beta$ is assumed to be sufficiently large while $H_R$ is sufficiently close to unity to make the quantum trajectories superluminal consistent with the
apparent superluminal OPERA neutrinos [\ref{bib:opera}].   As $q$ increases along the OPERA neutrino's quantum trajectory, the quantum factor $H_Q$ fluctuates
within the quantum factor's wedge between its maximum and minimum values.  For any $q \ge 0$, there exists a phase
shift  $-\pi < \varphi_{\mbox{\scriptsize min}} \le \pi$ for which $H_Q(E,\beta,\varphi;q)= H_{Q,\mbox{\scriptsize min}}$.  Hence, all values of $q$ are on some
superluminal quantum trajectory for the OPERA neutrino.  For any $q$, there also exists another phase shift  $-\pi < \varphi_{\mbox{\scriptsize max}} = \varphi_{\mbox{\scriptsize min}} \pm \pi \le \pi$ such that
the $H_Q(E,\beta,\varphi;q)= H_{Q,\mbox{\scriptsize max}}$, which implies the existence of also a subluminal quantum trajectory with a different phase
shift. And for an interior $q,t$-point within the wedge, there exist two crossing quantum trajectories.

For completeness, the non-relativistic dichromatic particle also exhibits theoretical superluminal behavior [\ref{bib:fp37a}].  The non-relativistic
dichromatic particle differs with the neutrino investigated herein by its non-relativistic wave number.  And elsewhere the relativistic particle deep
into the forbidden zone has been shown to be superluminal [\ref{bib:ijtp27}].  Also, if $\beta = \alpha = 2^{-1/2}$, then $\psi$ manifests a standing wave.

\begin{figure}
\centering
\includegraphics[scale=0.75]{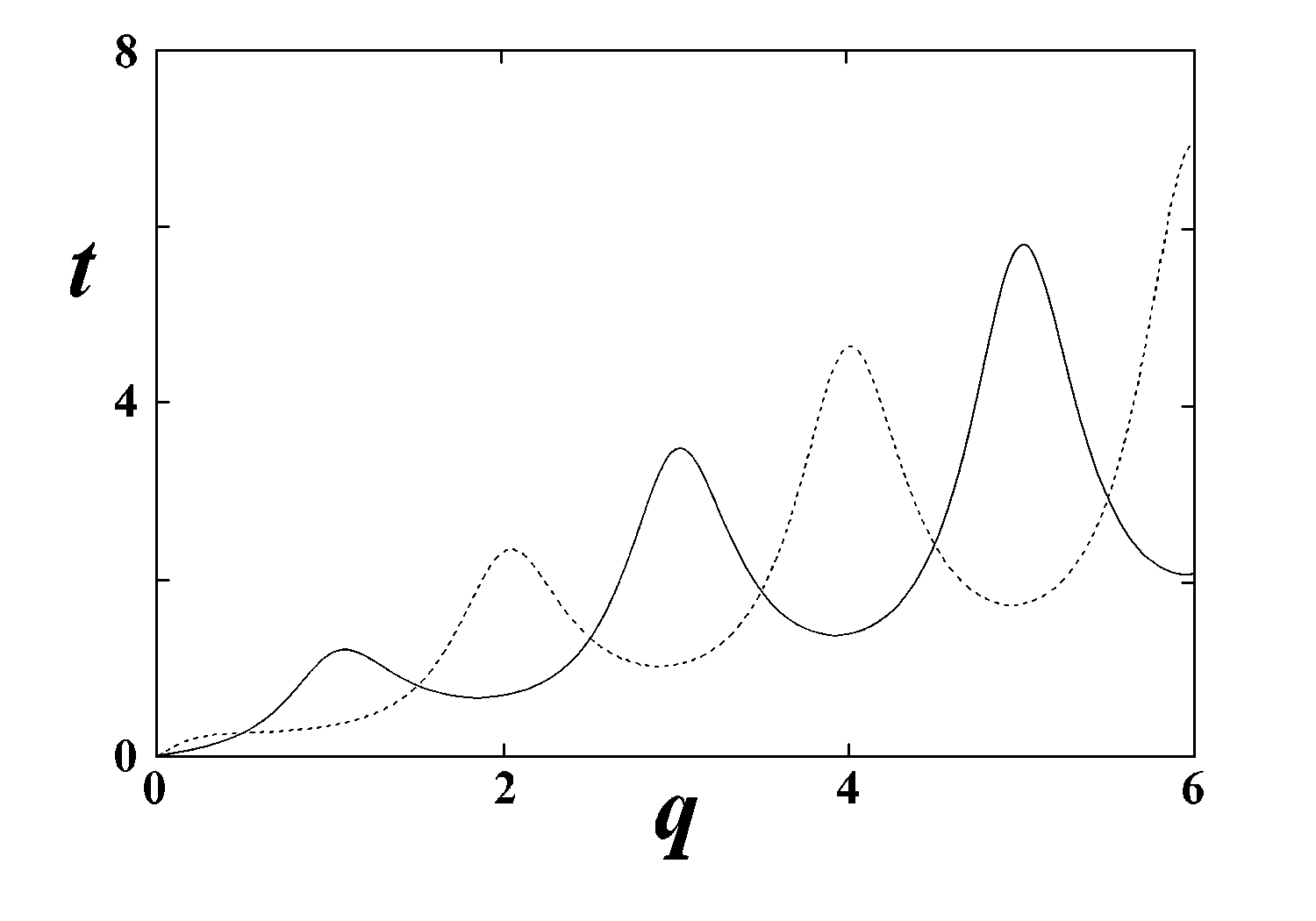}
\caption{\small Heuristic quantum trajectories in natural units ($\hbar,c,m=1,k=\pi/2$) with constants of the quantum motion $E=(\pi^2/4+1)^{1/2}$
(or alternatively $k=\pi/2$), $\beta = 0.28$, and $\varphi=0,\pi$.  The quantum trajectory for $\varphi=0$ is solid, while the quantum trajectory for $\varphi=\pi$
is dashed.}
\end{figure}

Figure 1 exhibits the quantum trajectories for an example where $\hbar,c,m=1$ (natural units) and where the constants of the quantum motion are given
by $E=(\pi^2/4+1)^{1/2}$ (or alternatively by $E$'s proxy $k=\pi/2$), $\beta = 0.28$, and $\varphi=0,\pi$.  While this $\beta$ is orders of magnitude
too large to be
representative of the OPERA anomaly, it does render a good heuristic example of quantum trajectories for dichromatic particles.  The quantum
trajectories have temporal turning points where the quantum trajectories alternately become temporally retrograde and forward. The self-entanglement
between the two spectral components, Eq.\ (\ref{eq:spectrum}), induces this retrograde motion.  In turn, retrograde motion induces nonlocality as exhibited on
Fig.\ 1 where the quantum trajectory permits multiple particle positions, $q$s, for selected times, $t$s, until a threshold time, $t_{\mbox{\scriptsize th}}$,
after which all times permit
multiple particle positions.  As $q$ increases, the turning points asymptotically approach the bounds prescribed for $H_Q$, Eqs.\ (\ref{eq:maxhq})
and (\ref{eq:minhq}).  The entwined quantum trajectories for $\varphi=0,\pi$ on Fig.\ 1 demonstrate the existence of concurrent superluminal and subluminal
propagation.  Note that there are two quasi-periodic meanderings for every wavelength as characteristic of nonlinear differential equations for
the interference effect due to the cosine term of Eq.\ (\ref{eq:eom}) has its frequency doubled, that is the $2kq$ component in the argument of the
cosine term has a span of $4\pi$ radians per wavelength.  For completeness, these quasi-periodic meanderings asymptotically approach
periodicity as $q \to \infty$ [\ref{bib:fp37a}].

While early latent, temporally retrograde segments may be suppressed as shown by Fig. 1, retrograde segments are realized as $q$ increases.  Any finite $\beta$ will
induce nonlocality for sufficiently large $q$ [\ref{bib:fp37a}]. This may be shown by investigating the equation of quantum motion, Eq.\ (\ref{eq:eom}). Temporal
turning points are smooth and exist where the reciprocal velocity, $dt/dq$, goes to zero. The behavior of reciprocal velocity may be derived
from Eq.\ (\ref{eq:eom}) and given by

\begin{equation}
\frac{dt}{dq} = \frac{d[q(c^{-1}H_R)]}{dq} \times H_Q\, + \, qc^{-1}H_R \times \frac{d(H_Q)}{dq}.
\label{eq:dtdq}
\end{equation}

\noindent The conditions for a temporal turning point, $dt/dq = 0$, may be simplified to

\begin{equation}
\frac{1}{q}\ +\ \frac{4\alpha \beta k \sin(2kq+\varphi)}{1+\alpha \beta \cos(2kq+\varphi)} = 0.
\label{eq:sdtdq}
\end{equation}

\noindent As $q$ may increase without bound in Eqs.\ (\ref{eq:dtdq}) and (\ref{eq:sdtdq}), the $1/q$ term in Eq.\ (\ref{eq:sdtdq}) will
decrease sufficiently to ensure the existence temporal turning points. The existence of temporally retrograde motion follows.

There is an alternative interpretation to segments of temporal retrograde motion: such retrograde motion is an antiparticle, here an
antineutrino, $\bar{\nu}$, that moves forward in time.  Under this interpretation, $\nu,\bar{\nu}$-pair creations occur at the temporal
turning point (local temporal minimum) associated with $H_{Q,{\mbox{\scriptsize min}}}$ while $\nu,\bar{\nu}$-pair annihilations
occur at the temporal turning point (local temporal maximum)
associated with $H_{Q,\mbox{\scriptsize max}}$. For nonrelativistic quantum trajectories, it has been shown for temporal turning points of
nonrelativistic quantum trajectories that neither pair creations need be an endoergic process nor pair annihilation need be an exoergic
process [\ref{bib:fp37a}].  Upon $\nu,\bar{\nu}$-pair creation at the local temporal minimum associated with $H_{Q,\mbox{\scriptsize min}}$, the $\bar{\nu}$
travels forward in time but spatially retrograde in the $-q$-direction while the $\nu$ travels forward in time and in the $+q$-direction.
This gives the quantum trajectory a bi-directional character manifesting internal backscatter.  The $\bar{\nu}$ travels in the $-q$-direction
to the local temporal maximum associated with $H_{Q,\mbox{\scriptsize max}}$.  There the $\bar{\nu}$ joins with another $\nu$, which has been
traveling on the preceding forward
segment toward the local temporal maximum, to form a $\nu,\bar{\nu}$-pair that is annihilated at that local temporal maximum.

The $\nu,\bar{\nu}$-pair creations may facilitate neutrino oscillations by possibly producing pairs of different flavors, $e,\mu,\tau,$ sterile, or,
perhaps a mixed pair thereof.  The bidirectional character of the quantum trajectory, discussed in the previous paragraph, also manifests neutrino
oscillation.   On the other hand, one using the $\psi$ representation would conclude that the wave packet would loose integrity
at a temporal turning point due to interference from backscatter.  If a quantum trajectory has retrograde segments, then there exists by Fig. 1,
as previously discussed, a threshold
time, $t_{\mbox{\scriptsize th}}$, for which the quantum trajectory is multi-valued with regard to $q$ for all $t>t_{\mbox{\scriptsize th}}$.  These $q$s for a
particular time $\check{t}>t_{\mbox{\scriptsize th}}$ form the set
$\{q_{\check{t}}\}$.  The smallest $q_{1,\check{t}} \in \{q_{\check{t}}\}$ is on a temporally forward segment of the quantum trajectory exhibiting a $\nu$ particle
of some particular flavor.  The next greater $q_{2,\check{t}} \in \{q_{\check{t}}\}$ is on a temporal retrograde segment of the quantum trajectory exhibiting a
$\bar{\nu}$ antiparticle of some flavor that may differ from the flavor of the $\nu$ particle associated with $q_{1,\check{t}}$.  This alternating pattern repeats
itself for the remainder of the quantum trajectory with the caveat that any rotation through the different flavors (neutrino oscillations) may be more arcane.
While the quantum trajectory for any $\check{t}$ may determine the entire $\{q_{\check{t}}\}$, any
measurement of some subset of $\{q_{\check{t}}\}$ would depend upon how the measuring experiment would be executed.
As the primary objective here is to investigate the OPERA superluminal anomaly,
it suffices that quantum trajectories for this investigation assume that no more than two types of neutrinos,
$\nu$ and $\bar{\nu}$, with the same flavor would participate.

\section{Application to OPERA experiment}

The OPERA collaboration observed [\ref{bib:opera}] with 6.2 $\sigma$ significance a relative superluminal deviation
$(v-c)/c = [2.37 \pm 0.32\ (\mbox{stat.})\ ^{+0.34}_{-0.24}\ (\mbox{sys.})] \times 10^{-5}$ for 17 GeV neutrinos.  From
Eqs.\ (\ref{eq:eom}) and (\ref{eq:minhq}), a selection of $\beta \approx 1.185 \times 10^{-5}$ is chosen to mimic the OPERA collaboration's neutrino
superluminal anomaly.  For 17 GeV neutrinos,
the mass of a particular flavor is considered to have a negligible effect smaller than $10^{-19}$ [\ref{bib:opera}].
For 17 GeV and a neutrino rest mass of 2 eV, the neutrino would have a wavelength given by $\lambda= hc (E^2-m^2c^4)^{-1/2}=72.932$ am.
The 730534.61$\pm$0.2 m between the CNGS target focal point at CERN and the origin of the OPERA detector
at LNGS equates to approximately $1.0016670729207\times10^{22}$  wavelengths.  The quantum trajectory for a simulated OPERA neutrino with
rest mass 2 eV and constants of the quantum motion $E=17$GeV, $\beta=1.185\times 10^{-5}$ and $\varphi=0$, is investigated herein.
Segments of its computed quantum
trajectory are presented in Figs. 2(a), 2(b) and 2(c).

\begin{figure}
\centering
\includegraphics{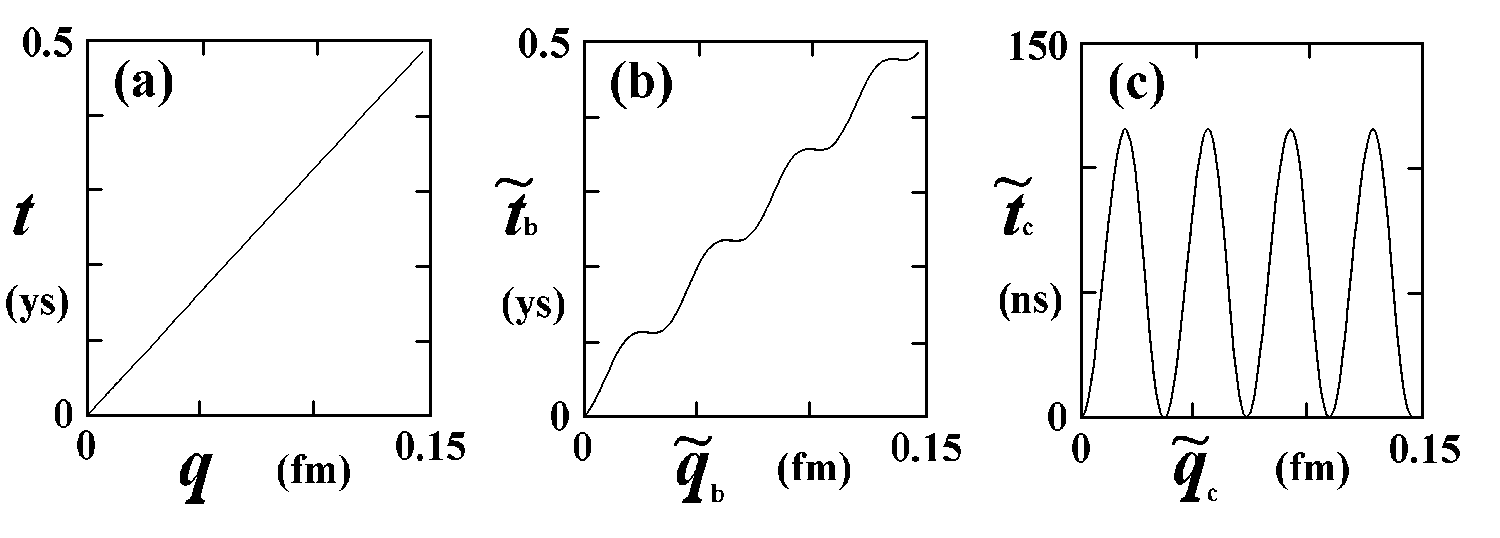}
\caption{\small The initial 2 wavelengths ($\lambda$) are exhibited on 2(a) for a theoretical quantum trajectory with sufficient
backscatter to simulate the OPERA superluminal anomaly. On 2(b) is a 2 $\lambda$ section of a theoretical quantum
trajectory that simulates the OPERA superluminal anomaly in the region where the transition to temporal
retrograde motion begins. The displayed origin of this section for 2(b) is at 3700 wavelengths ($q = 269.8$ pm)
and 900.1 ys. Hence, the coordinates for 2(b) are $\tilde{q}_{\mbox{\scriptsize b}} =q-269.8$ pm and $\tilde{t}_{\mbox{\scriptsize b}}=t-900.1$ ys.
On 2(c) is a 2 $\lambda$
section of a theoretical quantum trajectory with sufficient backscatter to simulate the OPERA superluminal
anomaly at the OPERA detector. The displayed origin of this section for 2(c) is at 730534.61 m and 2.437 ms.
Hence, the coordinates for 2(c) are $\tilde{q}_{\mbox{\scriptsize c}}=q-730534.61$ m and $\tilde{t}_{\mbox{\scriptsize c}}=t-2.437$ ms.
The time scale for 2(c) differs with those of 2(a) and 2(b).}
\end{figure}

Figure 2(a) shows the quantum trajectory for the first 2 wavelengths. To the eye, it appears as a straight line, but it does have some microscopic
quasi-periodic meanders within the $H_Q$ wedge that are insufficient to cause time reversals.  Figure 2(b) exhibits the quantum trajectory for 2 wavelengths
beginning 3700 wavelengths ($q=269.8$ fm) from the CNGS target focal point and in the transition region where time reversals begin to appear.
The quasi-periodic meanderings
of the quantum trajectory are apparent in Fig.\ 2(b).  The duration of one wavelength (two meanderings) is 0.2433 ys as exhibited on Fig.\ 2(b).  The time coordinate (vertical axis) of Fig.\ 2(b) is $\tilde{t}_{\mbox{\scriptsize b}}=t-900.4$ ys
while the distance coordinate (horizontal axis) is $\tilde{q}_{\mbox{\scriptsize b}}=q-269.8$ fm.
Figure 2(c) shows the quantum trajectory
in the approximate vicinity of the OPERA detector at LNGS, which is $1.00166707\times10^{22}$ wavelengths (730534.61 m) from the CNGS target focal point
at CERN.  The computed time of flight, $TOF_{\nu}$, for a neutrino to reach the the lower (minimum) caustic at the OPERA detector at LNGS is 2.437 ms.
Segments of the quantum trajectory alternate between forward and retrograde motion with regard to time.  Between segments, there exist
smooth turning points near the upper and lower caustics where neutrino speed becomes instantaneously infinite.  [Note that this instantaneously infinite
speed is consistent with $dt/dq \to 0$, cf.\ Eqs.\ (\ref{eq:dtdq}) and (\ref{eq:sdtdq}).]  Nevertheless, the velocity remains integrable as
substantiated by Fig.\ 2(c).  The coordinates for 2(c) are
$\tilde{q}_{\mbox{\scriptsize c}}=q-730534.61$ m and $\tilde{t}_{\mbox{\scriptsize c}}=t-2.437$ ms.   Note that the time scale of Fig.\ 2(c) is much
coarser than those of Figs.\ 2(a) and 2(b) to accommodate the size of the durations
of the temporally forward and retrograde segments.  These durations increase as $q$ increases.  While not apparent in Fig.\ 2(c), corresponding points on successive wavelengths advance in
time about 0.2433 ys consistent with Figs.\ 2(a) and 2(b).  The $TOF_{\nu}$ computation predicts an early arrival time
with respect to that assuming the speed of light in a vacuum, $TOF_c$. The time difference is given by

\begin{equation}
\delta t = TOF_c - TOF_{\nu} = 57.75\ \mbox{ns}
\label{eq:delt}
\end{equation}

\noindent while OPERA observed $\delta t = (57.8\pm7.8(\mbox{stat.})^{+8.3}_{-5.9}(\mbox{sys.}))\ \mbox{ns}$. The relative difference in the computed
neutrino velocity, $v$, and $c$ is by Eq.\ (\ref{eq:delt})

\begin{equation}
(v-c)/c = 2.37\times10^{-5}
\label{eq:rvdif}
\end{equation}

\noindent while OPERA observed $(v-c)/c=(2.37\pm0.32(\mbox{stat.})^{+0.34}_{-0.24}(\mbox{sys.}))\times10^{-5}$.  But the constants of
quantum motion, $\beta=1.185\times 10^{-5}$ and $\varphi=0$, where chosen to match the OPERA anomaly for a 17 GeV neutrino at $q=730534.61$ m.  Note
that value of $\beta$ used here is well within the reported value [\ref{bib:opera}] for $\bar{\nu}_{\mu}$ contamination at the OPERA detector of
less than 2.1\%.  Again, there exists for other points, $q$, on the quantum trajectory for which some value of the constant of the quantum motion $\varphi$ in the
range $(-\pi,+\pi)$ would also render Eq.\ (\ref{eq:rvdif}) true.

We note in Fig.\ 2(c) that the time duration of the retrograde segments has increased to about 115.5 ns.  These durations will continue to increase with
increasing $q$ as the quantum trajectory proceeds out the wedge of $H_Q$.  As discussed earlier, a temporally retrograde segment may be considered to manifest an
antineutrino, $\bar{\nu}$, traveling forward in time but in a negative $q$-direction.  While all segments of the quantum trajectories between
$\nu,\bar{\nu}$-pair creations in the vicinity of $H_{Q,\mbox{\scriptsize min}}$ and $\nu,\bar{\nu}$-pair annihilations in the
vicinity of $H_{Q,\mbox{\scriptsize max}}$ are spatially one-quarter wavelength long (18.233 am) and temporally forward with an approximate
duration of 115.5 ns as previously noted.  The average speed over these quarter-wavelength segments is quite subluminal, not even pedestrian, at
approximately 0.1579 nm/s.  Hence, neither $\nu$ particle advancing in the positive $q$-direction nor the $\bar{\nu}$ antiparticle receding in the
negative $q$-direction (spatial retrograde motion) are tachyons in neighborhoods that do not include temporal turning points.  As such, the subluminal
portions of neutrino's quantum trajectory are not subject to the restriction of Cohen and Glashow [\ref{bib:cg}] that tachyons could not sustain
superluminal quantum motion due to a loss of energy to spontaneous $e^-,e^+$-pair creations.  On the other hand, the quantum trajectories are
superluminal at isolated temporal points where either $\nu,\bar{\nu}$-pair creations or $\nu,\bar{\nu}$-pair annihilations
exist.  These $\nu,\bar{\nu}$-pair creations or $\nu,\bar{\nu}$-pair annihilations are respectively neither endoergic nor exoergic as energy $E$
is explicitly one of the constants of quantum motion for the quantum trajectory and so used in Jacobi's theorem to develop the relativistic equation
of quantum motion, Eq.\ (\ref{eq:eom}). Thus, the restrictions of Cohen and Glashow [\ref{bib:cg}] are also not applicable to the superluminal
neighborhood associated with temporal turning points.  The findings of this paragraph may be generalized for the entire quantum trajectory.  And the
full quantum trajectory through its nonlocality may have superluminal transit time even if all of its individual segments may have subluminal average absolute values of speed over their respective segments.

Had the 2007 MINOS neutrino velocity anomaly [\ref{bib:minos}],

\[
\frac{(v_{\mbox{\scriptsize \scshape minos}}-c)}{c}=[5.1\pm2.9(\mbox{stat.}+\mbox{sys.}] \times 10^{-5} \ \mbox{68\% confidence limits,}
\]

\noindent been chosen for investigation, then the constant of quantum motion $\beta_{\mbox{\scriptsize \scshape minos}} = 2.55 \times 10^{-5}$ would have been selected.  It is noted that
the MINOS collaboration never asserted that MINOS neutrinos were superluminal as the MINOS confidence limits were insufficient.

As $H_{Q,{\mbox{\scriptsize max}}}$ and $H_{Q,{\mbox{\scriptsize min}}}$ are just functions of $\beta$ as already noted and as $H_Q$ is the more dominant
than $H_R$ for neutrinos at the energies of the OPERA experiment, the relative difference $(v-c)/c$ is more a function of the constant of the quantum motion $\beta$
than a function of the constant of the quantum motion $E$.  This is consistent with the OPERA experiment where any energy dependence of $v$ is less than one $\sigma$
 over the energy range of the experiment [\ref{bib:opera}].

However, OPERA collaboration obtained $\delta t$ by comparing the temporal distributions of neutrino interactions at LNGS with the temporal distribution of protons hitting
the CNGS target [\ref{bib:opera}].  This implies that many interactions over the range $(-\pi,+\pi)$ for the constant of the quantum motion $\varphi$ must be
considered.  The average for the quantum factor $H_Q$ of Eq.\ (\ref{eq:eom}) is given by averaging over all quantum trajectories, each specified by
its $\varphi$, that may intercept a given $q$.  Averaging over $\varphi$ is given by [\ref{bib:dwight}]

\[
<H_Q>_{\varphi} = \frac{1}{2\pi} \int^{+\pi}_{-\pi} H_Q\,d\varphi = \frac{(\alpha^2-\beta^2)}{2\pi} \int^{+\pi}_{-\pi} \frac{d\varphi}{1 + 2 \alpha \beta
\cos(2kq+\varphi)} = \frac{(\alpha^2-\beta^2)}{2\pi} \frac{2\pi}{[(\alpha^2-\beta^2)^2]^{1/2}} = 1.
\]

\noindent  So, uniform $\varphi$-averaging over all quantum trajectories that may intercept $q$ does not render an aggregate superluminal motion. This result is consistent
with the results of Matone [\ref{bib:m}] and Faraggi [\ref{bib:f}] that while particular quantum trajectories may have superluminal transit times, the
quantum correction (here $H_Q$) does not necessarily support the existence of superluminal motion for the general case.

Possible sources of backscatter in the quantum reduced action or SKGE's $\psi$ include, but not limited to, the CNGS neutrino beam assembly,
neutrino oscillations, transmission effects through the earth including the Mikheyev-Smirnov-Wolfenstein (MSW) effect [\ref{bib:msw}], thermal
effects [\ref{bib:m2}] and interference due to a quantum analogy to reverberation, and the OPERA detector at LNGS.  If any single cause or some
combination thereof systematically introduces a
biased $\varphi$-distribution, perhaps in combination with a biased $\beta$ distribution, then this could permit superluminal quantum motion.  Then
the observed OPERA superluminal quantum motion [\ref{bib:opera}] for could be attributed to an internal interference effects described by particular values
of the constants of quantum motion $\beta$ and $\varphi$.  We shall briefly study three cases that lead to superluminal neutrinos

 In the first case, we consider $\nu,\bar{\nu}$-pair creations.  Neutrino oscillations that create $\nu,\bar{\nu}$-pairs are adduced as a candidate source for backscatter with a favorable biased $\varphi$-distribution.  These $\nu,\bar{\nu}$-pair creations occur at local minima temporal turning points as shown on Figs.\ 1 and 2(c).  Pair creation phenomenon at maximum reinforcement by coherent interference is theoretically substantiated by Refs.\ \ref{bib:fp37a} and \ref{bib:fp37b}.  If the OPERA experiment systematically only detected these pair creations, then a $\beta=1.185\times 10^{-5}$, as used for Eq.\ (\ref{eq:rvdif}), would explain OPERA superluminal propagation. The OPERA experiment would systematically select $\varphi_{\mbox{\scriptsize min}}$ to detect $\nu,\bar{\nu}$-pair creation.

On the other hand, let us suppose for a second case that the OPERA experiment could detect not only $\nu,\bar{\nu}$-pair creations but also $\nu,\bar{\nu}$-pair annihilations as well.  The forward and backscatter spectral components of $\psi$ are in coherent reinforcement at the local minimum temporal turning points [\ref{bib:fp37a},\ref{bib:fp37b}] where $\nu,\bar{\nu}$-pairs are created.  For coherent reinforcement, the relative amplitude of the eigenfunction, $\psi_+$, would be proportional to $\alpha+\beta$ by Eq.\ (\ref{eq:spectrum}) where the spectral components would be in phase. For $\nu,\bar{\nu}$-pair annihilations at local maximum temporal turning points, the relative amplitude of the eigenfunction, $\psi_-$ would be $\alpha - \beta$ where the spectral components would be out of phase [\ref{bib:fp37a},\ref{bib:fp37b}].  The predicted relative difference of the neutrino velocity, analogous to Eq.\ (\ref{eq:rvdif}), for detecting just both $\nu,\bar{\nu}$-pair creations and annihilations would have the form

\begin{equation}
1\ -\ \overbrace{\underbrace{\frac{\alpha-\beta}{\alpha+\beta}}_{H_{Q,\mbox{\scriptsize min}}}\, \times\, \underbrace{(\alpha+\beta)^2}_{\psi_+^{\dagger}\psi_+^{\ }}}^{\mbox{pair creation}}\,\div \, 2 \ +\ \overbrace{\underbrace{\frac{\alpha+\beta}{\alpha-\beta}}_{H_{Q,\mbox{\scriptsize max}}}\, \times\, \underbrace{(\alpha-\beta)^2}_{\psi_-^{\dagger}\psi_-^{\ }}}^{\mbox{pair annihilation}}\, \div \, 2\  =\ 2\beta^2\ =\ 2.37\, \times\, 10^{-5}
\label{eq:cran}
\end{equation}

\noindent  where $\psi^{\dagger}\psi$ is the Born weighting function.  The divisor ``2" in Eq.\ (\ref{eq:cran}) is the normalization $\psi_+^{\dagger}\psi_+^{\ }+\psi_-^{\dagger}\psi_-^{\ }=2$ in Eq.\ (\ref{eq:spectrum}).  As the internal interference between spectral components of $\psi$ are in reinforcement at local time minima while in opposition at local time maxima, the Born weighting function $\psi^{\dagger}\psi$ renders preponderant weighting to minimum transit times  ---  hence, superluminal neutrinos.  The constant of the motion $\beta$ that would be consistent with OPERA results for detecting just both $\nu,\bar{\nu}$-pair creations and annihilations is given by $\beta = 3.442 \times 10^{-3}$.  This value of $\beta$ is also well within the reported value [\ref{bib:opera}] for $\bar{\nu}_{\mu}$ contamination at the OPERA detector of less than 2.1\%.  Note that $\beta$ for this second case must be greater than the $\beta$ for the first case as $\nu,\bar{\nu}$-pair annihilations by themselves would create subluminal neutrinos.

In the third case, the neutrino may be detected at any point $q$ within the OPERA detector at LNGS along the neutrino's trajectory for any value of the constant of the motion $\varphi$.  The cross section of neutrino detection at $q$ for each trajectory with a constant of the motion $\varphi$ has a Born weighting function given by $\psi^{\dagger}(q)\psi(q) = 1 + 2\alpha\beta\cos(2kq+\varphi)$ from Eq.\ (\ref{eq:spectrum}).  As the OPERA detector is more than $10^{22}$ wavelengths down stream from the CNGS target focal point at CERN, evaluation of the relative difference of neutrino velocity, Eq.\ (\ref{eq:rvdif}) may be simplified to considering the contributions of only $H_q$.  We average $H_q$ weighted by $\psi^{\dagger}(q)\psi(q)$ over $\varphi$ to account for all trajectories that intercept $q$.  The predicted relative difference of the neutrino velocity, analogous to Eq.\ (\ref{eq:rvdif}), for detecting a neutrino weighted by $\psi^{\dagger}(q)\psi(q)$ has the form

\[
1 - \frac{\int_{-\pi}^{\pi} H_Q \psi^{\dagger}\psi \,  d\varphi}{\int_{-\pi}^{\pi} \psi^{\dagger}\psi \, d\varphi}\ =\ 2\beta^2\ =\ 2.37\, \times\, 10^{-5}.
\]

\noindent  Analogous to the second case, the interference between the spectral components of $\psi$ are in reinforcement if $H_Q < 1$ and in opposition if $H_Q > 1$.  This correlation  between the Born weighting function $\psi^{\dagger}\psi$ and $H_Q$ is due to their common dependence upon the term $\cos(2kq+\varphi)$ in Eqs.\ (\ref{eq:spectrum}) and (\ref{eq:eom}). As $\int_{-\pi}^{\pi} \cos(kq+\varphi) \, d\varphi = 0$, superluminal times follow for neutrinos with internal backscatter.   The value of the constant of the motion $\beta$ that would be consistent with OPERA results is $\beta  = 3.442 \times 10^{-3}$.  This value for $\beta$ is the same as the value for the second case. This sameness may be explained by

\[
\psi^{\dagger}\Big|_{\varphi=\breve{\varphi}}\psi\Big|_{\varphi=\breve{\varphi}} + \psi^{\dagger}\Big|_{\varphi=\breve{\varphi}+\pi}\psi\Big|_{\varphi=\breve{\varphi}+\pi} = 2
\]

\noindent where the difference of $\pi$ between the values for the constant of motion $\varphi$ in the two weighting functions causes their interference contributions to cancel each other.

\section{Conclusion}

In conclusion, quantum trajectories with backscatter can explain superluminal propagation for quantum trajectories of neutrinos.  Backscatter can be incorporated into a general solution for the quantum reduced action so that self-entanglement is maintained in the generator of quantum motion of neutrinos.  Backscatter may also be incorporated into an eigensolution of the associated Schr\"{o}dinger equation so that self-entanglement maintains coherent interference.   Entanglement, Eq.\ (\ref{eq:spectrum}), begets nonlocality that in turn begets overall superluminal transit times for the full trajectory even if transit times for all segments of the trajectory (the absolute value of the time to transit from a temporal extremum to the next temporal extremum) are subluminal.  These segments alternate between being forward or retrograde in time.  For any point on the quantum trajectory with constants of quantum motion
energy, $E$, and finite backscatter, $\beta > 0$, there exists a third constant of quantum motion phase shift, $\varphi$, for which the quantum trajectory is superluminal.  When the interference due to backscatter is averaged for a uniform distribution of phase shift, $\varphi$, then no superluminal propagation in general is expected.  If the OPERA experiment in a systematic way favorably biases the interference due to backscatter, then superluminal propagation may be expected. Different physically inspired processes that systematically favorably bias the interference are examined.  Processes that include the favorable correlation with regard to phase shift, $\varphi$, between the Born weighting function $\psi^{\dagger}\psi$ and quantum factor $H_Q$ in the equation of motion, Eq.\ (\ref{eq:eom}), lead to theoretical superluminal neutrinos.  These superluminal neutrinos can replicate the apparent superluminal OPERA observations by adjusting the degree of backscatter, $\beta$, a constant of the motion.  The degree of superluminality beyond a threshold $\beta$ increases as $\beta$ increases.

The restrictions of Cohen and Glashow [\ref{bib:cg}] are applicable to neither subluminal nor superluminal portions of the quantum trajectories for which energy is a constant of the quantum motion.  At the temporal turning points of a quantum trajectory where the velocity of the quantum trajectory becomes instantaneously infinite, either $\nu,\bar{\nu}$-pair creations or annihilations occur instead of Cohen and Glashow $e^+,e^-$-pair creations.

A byproduct of these investigation is insight into neutrino oscillation.  The $\nu,\bar{\nu}$-pair creations occur at the local time minima of the quantum trajectory while $\nu,\bar{\nu}$-pair annihilations occur at local time maxima.

\bigskip

\noindent ACKNOWLEDGEMENT:  I heartily thank M.\ Matone, A.\ E.\ Faraggi and R.\ Carroll for their continued assistance and encouragement through the last 15
years in my endeavors in quantum trajectories.

\bigskip

\noindent REFERENCES

\begin{enumerate}\itemsep -.06in

\item \label{bib:opera} OPERA collaboration: T.\ Adams {\itshape et al}, ``Measurements of the neutrino velocity with the OPERA detector in the CNGS beam",
[arXiv:1109.4897v2].

\item \label{bib:m} M.\ Matone, ``Superluminal neutrinos and a curious phenomenon in the relativistic Hamilton-Jacobi equation", [arXiv:1109.6631v2].

\item \label{bib:m2} M. Matone, ``Neutrino speed and temperature", [arXiv:1111.0270v3]

\item \label{bib:f} A.\ E.\ Faraggi, ``OPERA data and the equivalence postulate of quantum mechanics", [arXiv:1110.1857v2].

\item \label{bib:pl450} A.\ E. Faraggi and M.\ Matone, Phys.\ Lett.\ {\bfseries B 450}, 34 (1999), [hep-th/9705108].

\item \label{bib:fm} A.\ Faraggi and M.\ Matone, Int.\ J.\ Mod.\ Phys.\ {\bfseries A 15} 1869 (2000), [hep-th/9809127].

\item \label{bib:bfm} G.\ Bertoldi, A.\ E.\ Faraggi and M.\ Matone, Class.\ Quant.\ Grav.\ {\bfseries 17}, 3965 (2000), [hep-th/9909201].

\item \label{bib:mb} A.\ Mecozzi and M.\ Bellini, ``Superluminal group velocity of neutrinos", [arXiv:1110.1253v2].

\item \label{bib:ikmg} D.\ Indumathi, R.\ K.\ Kaul, M.\ V.\ N,\ Murthy, and G.\ Rajasekaran, ``Group velocity of neutrino waves", [arXiv:1110.5453v2].

\item \label{bib:bbps} M.\ V.\ Berry, N.\ Brunner, S.\ Popescu, and P.\ Shukla, ``Can apparent superluminal neutrino speeds be explained as a quantum
weak measurement?", [arXiv:1110.2832v2].

\item \label{bib:tm} T.\ Morris, ``Superluminal group velocity through near-maximal neutrino oscillations", [arXiv:1110.2463v3]

\item \label{bib:fp37a} E.\ R.\ Floyd, Found.\ Phys.\ {\bfseries 37}, 1386 (2007), [quant-ph/0605120v3].

\item \label{bib:ijtp27} E.\ R.\ Floyd, Int.\ J.\ Theor.\ Phys.\ {\bfseries 27}, 273 (1988).

\item \label{bib:prd29} E.\ R.\ Floyd, Phys.\ Rev.\ {\bfseries D 29}, 1842 (1984).

\item \label{bib:prd34} E.\ R.\ Floyd, Phys.\ Rev.\ {\bfseries D 34}, 3246 (1986).

\item \label{bib:hm} C.\ E.\ Hecht and J.\ E. Mayer, Phys.\ Rev.\ {\bfseries 106}, 1156 (1953).

\item \label{bib:pe5} E.\ R.\ Floyd, Phys.\ Essays\ {\bfseries 5}, 130 (1992).

\item \label{bib:milne} W.\ E.\ Milne, Phys.\ Rev.\ {\bfseries 35}, 863 (1930).

\item \label{bib:fpl9} E.\ R.\ Floyd, {Found. Phys. Lett.} {\bf 9}, 489 (1996), [quant-ph/9707051].

\item \label{bib:rc} R.\ Carroll, J.\ Can.\ Phys. {\bfseries 77}, 319–325 (1999), [quant-ph/9903081].

\item \label{bib:cg} A.\ G.\ Cohen and S.\ L.\ Glashow, Phys.\ Rev.\ Lett.\ {\bfseries 107}:181803 (2011), [arXiv:1109.6562].

\item \label{bib:minos} MINOS collaboration: P.\ Adamson {\itshape et al}, Phys.\ Rev.\ {\bfseries D 76}, 072005 (2007), [arXiv:0706.0437v3].

\item \label{bib:dwight} H.\ B.\ Dwight, {\itshape Tables of Integrals and Other Mathematical Data} 4th ed.\ (MacMillan,
New York, 1961) \P 858.520, \P 858.521, \P 858.524, and \P 858.525.

\item \label{bib:msw} L.\ Wolfenstein, Phys.\ Rev.\ {\bfseries D 17}, 2369 (1978).

\item \label{bib:fp37b}  E.\ R.\ Floyd, Found.\ Phys.\ {\bfseries 37}, 1403 (2007), [quant-ph/0605121v3]

\end{enumerate}

\end{document}